\documentclass[jltp, 12pt, preprint]{revtex4-1}
\usepackage{graphicx}
\begin{document}
\title{Parallel Critical Field in Thin Niobium Films:  Comparison to Theory} 
\author{P.R. Broussard}
\email{phill.broussard@covenant.edu}
\affiliation{Covenant College, Lookout Mountain, GA 30750}


\begin{abstract}
For the first time, a comparison to the predicted behavior for parallel critical field is carried out for the model of Kogan and the model of Hara and Nagai.  In this study, thin niobium films in the moderately dirty regime were considered.  Experimental values of the $-C_{2}$ term are seen to be  lower than those from the model of Hara and Nagai.  A possible reason for this could be not including the non spherical Fermi surface of niobium into the model.  There is clearly disagreement with the model of Kogan as the films get cleaner and thinner, and two films which should be below his critical thickness still show positive values of  $-C_{2}$, in disagreement with his theory.   
\end{abstract}

\maketitle

\section{Introduction}
The study of the parallel upper critical field in thin films continues to be an area of research as seen in the work on amorphous Pb films\cite{Gardner} and a S/F bilayer,\cite{Lenk} both of which discussed the prediction by V. Kogan on a possible enhancement of superconductivity in thin clean superconducting films in applied parallel magnetic fields.\cite{Kogan,Kogan2}  This prediction was specifically mentioned in the theoretical work by Scotto and Pesch\cite{SP} and later by Hara and Nagai\cite{HN} on thin superconducting films, where they claimed there should not be any such enhancement.  Hara and Nagai in particular made predictions for the parallel critical field slope near the zero field transition temperature for thin superconducting films as a function of film thickness and resistivity, but this result has not yet been experimentally verified.  In this work the experimental values of  the parallel critical field slope near the zero field transition temperature are compared to the predictions of Kogan as well as Hara and Nagai for superconducting niobium films in the moderately dirty limit,  particularly for two films thinner than their critical thickness predicted by Kogan.

\section{Theoretical Considerations}
The first prediction for a superconducting film under a magnetic field applied along the film surface when the film is in the 2D limit was done by Tinkham,\cite{Tinkham} where for a dirty film (to be explained below) with thickness $d<1.84\xi_{\mathrm{GL}}(T)$, where $\xi_{\mathrm{GL}}(T)$ is the Ginzburg-Landau coherence length, he predicted that the parallel critical field as a function of temperature would follow the equation
\begin{equation}
\label{Tink1}
B_{c2||}(T)=\frac{\phi_{0}\sqrt{3}}{\pi\xi_{\mathrm{GL}}(T)d},
\end{equation}
where $\phi_{0}=\frac{h}{2e}$ is the flux constant (in MKS units).  This equation would predict that instead of $B_{c2}\propto(1-t)$ one would see instead $B^{2}_{c2}\propto(1-t)$, where $t$ is the reduced temperature.  In addition there would be a definite angular dependence of the critical field, given by 
\begin{equation}
\label{Eq6}
\big\{\frac{B_{c2}(T,\theta)|\cos(\theta)|}{B_{c2\bot}(T)}\big\}+\big\{\frac{B_{c2}(T,\theta)\sin(\theta)}{B_{c2||}(T)}\big\}^{2}=1,
\end{equation}
where $\theta$ is the angle between the applied magnetic field and the sample normal.

V. Kogan\cite{Kogan} and in a later paper with N. Nakagawa\cite{Kogan2} extended the above calculation by considering cleaner films, or arbitrary values of $\lambda_{tr}= \xi/\ell$  where $\ell$ is the elastic mean free path, $\xi=\hbar v_{F}/(2\pi k_{B}T_{c0})$, $T_{c0}$ is the zero field critical temperature and $v_{F}$ is the Fermi velocity in the material.  A dirty material has $\lambda_{tr}$ very large ($>>1$) and a clean material has $\lambda_{tr}$ very small ($<<1$).  Kogan predicted that Eq. \ref{Tink1} would be modified as
\begin{equation}
\label{Kogan1}
B_{c2||}(T)=\frac{\phi_{0}\sqrt{3}}{\pi\xi_{\mathrm{GL}}(T)\sqrt{d^{2}-d^{2}_{c}}},
\end{equation}
where $d_{c}$ is the critical thickness, defined by 
\begin{equation}
\label{Eq1}
d_{c}=\sqrt{7.2\gamma(\lambda_{\mathrm{tr}})\ell^2} ,
\end{equation}
and $\gamma(x)$ is given by
\begin{equation}
\label{Eq2}
\gamma(x)=\frac{x^{2}\sum_{n=0}^{\infty}(2n+1)^{-2}(2n+1+x)^{-3}}{\sum_{n=0}^{\infty}(2n+1)^{-2}(2n+1+x)^{-1}}.
\end{equation}
Kogan claimed that an enhancement in the $T_{c}$ would take place for parallel field for films less than the critical thickness, $d<d_{c}$.  

Scotto and Pesch's work\cite{SP} carefully considered the parallel upper critical field for films with varying thicknesses and mean free paths.  Their work was primarily a numerical study describing the parallel critical field over all temperatures, film thicknesses, and values of $\lambda_{tr}$, but showed that the enhancement of the critical temperature claimed by Kogan is not consistent with their model.  Hara and Nagai\cite{HN} working with the theory of Scotto and Pesch, considered the slope of the parallel critical field near the zero field critical temperature.
From their work the parallel critical field near $T_{c0}$ can be written in closed form as
\begin{equation}
\label{Eq3}
t_{C}(\lambda)=1+C_{2}\lambda^{2}+\cdots ,
\end{equation} 
where $t_{C}=T_{c}(B)/T_{c0}$, $\lambda=(eB/\hbar)\xi^{2}$ and 
\begin{equation}
\label{Eq4}
C_{2}=\sum_{n=0}^{\infty}\Big\{\frac{8}{15 \widetilde{\epsilon_{n}}^{3}\epsilon_{n}^{2} }-\frac{2D^{2}}{9 \widetilde{\epsilon_{n}}\epsilon_{n}^{2} }-\frac{8}{\widetilde{\epsilon_{n}}^{4}\epsilon_{n}^{2}D}\int_{0}^{\pi/2} \sin^{3}(\theta)\cos^{3}(\theta)\tanh\big(\frac{{\widetilde{\epsilon_{n}}} D}{2 \cos(\theta)}\big)d\theta \Big\} ,
\end{equation}
where 
 $\epsilon_{n}=2n+1$, $ \widetilde{\epsilon_{n}}=2n+1+1/\lambda_{tr}$, 
and $D=d/\xi$.   Hara and Nagai showed that $C_{2}$ is negative definite for all values of film thicknesses and mean free paths, in clear contradiction to Kogan's claim.

To clearly show the difference, the equation by Kogan can be recast into the language of Hara and Nagai.  Using Eq. \ref{Kogan1} and the equations for $\xi_{\mathrm{GL}}(T)$ in the work by Orlando {\it et al.}\cite{Orlando}, Kogan's work can be cast in the same form as Eq. \ref{Eq3}, with his value of $C_{2}$  given by the following expression:
\begin{equation}
\label{Eq5}
C_{2,\mathrm{Kogan}}=\frac{7\zeta(3)}{36}\chi(\lambda_{tr})(D_{c}^{2}-D^{2})
\end{equation}
where $\zeta(n)$ is the Riemann zeta function, $\chi(x)$ is the Gor'kov function, given as in the work by Orlando {\it et al.}\cite{Orlando}, $D=d/\xi$, and $D_{c}=d_{c}/\xi$ with $d_{c}$ given in Eq. \ref{Eq1}.  Clearly, Kogan predicts that for films thinner than the critical thickness, the value of $C_{2}$ will change sign, leading to, as Kogan claims, an enhancement of the critical temperature with applied parallel field.

One extra complication that has to be addressed is the strong coupling correction.  Theories built on the work of BCS/GLAG models of superconductivity do not predict measured values for some superconducting materials correctly due to the issues of strong coupling, that are not included in those weak coupled models.  As shown by Orlando {\it et al.}\cite{Orlando} one can correct for this by simply multiplying the theoretical value of perpendicular upper critical field by the corresponding strong coupling correction factor.  Since niobium is strong coupled, this correction must be included.  However there is no correction factor given for parallel critical field.  Since from the work by Orlando it is known that $\eta_{\mathrm{Bc2\perp}}=\eta_{\xi_{\mathrm{GL}}}^{-2}$, and since in Eq. \ref{Tink1} there is one power of $\xi_{\mathrm{GL}}$, it would seem prudent to use as the correction factor for $B_{c2||}$ the term $\sqrt{\eta_{\mathrm{Bc2\perp}}}$.

With all the theoretical exploration of this topic, it was hoped that comparison to actual measurements might be made to see whether the model  of Kogan's or the Scott/Pesch/Hara/Nagai model best fits the data.

\section{Film Production}
The niobium films were produced by magnetron sputtering onto substrates from a 99.95\% pure Nb target in UHP Ar at a pressure of 133 mPa.  Base pressures in the system were typically 4 $\mu$Pa, with LN$_{2}$ cooled surfaces near the substrates to remove water vapor.  Films were typically grown at either room temperature or with the substrate stage heated from behind by quartz lamps to a temperature of 300 C, as measured behind the substrate stage. Measurements of the substrate temperature show they are typically at 230 C under these conditions.  Growth rates were typically 8 nm/min, with the samples rotated during deposition to improve uniformity both from a thermal and deposition viewpoint.  Substrates were either (100) Silicon or A-plane and C-plane sapphire.   The substrates were either used as provided or cleaned with a RCA-1 process.\cite{RCA}   Film thicknesses were based off reference samples that were measured by stylus profilometer at the IEN Lab in Georgia Tech and a 5\% uncertainty to those values has been assigned.  Film thicknesses in this study varied from 16-52 nm.  Table \ref{Table} gives a listing of the samples used in this study.

\section{Transport measurements}
All samples had their resistivity measured by the Van der Pauw technique.\cite{VdP}  The samples were mounted on an OFHC copper platform cooled by a cryocooler down to a base temperature of $\approx$ 6 K.  A magnetic field up to $\approx$ 0.7 T could be applied to the sample, and the angle of the field could be varied from perpendicular to parallel to the film surface.

Upper critical field measurements were made at constant field, with the temperature swept from above to below the transition, typically at 0.1-0.2 K/min.  The critical temperature for the applied field was defined by the midpoint of the transition.  Field orientation to the sample was determined by measuring the transition at constant temperature and sweeping field up for various angles to find the maximum field and therefore determine the parallel field orientation.  

In order to compare to the predictions of Hara and Nagai, measured values of $d$, $T_{c0}$ and $\rho_{\mathrm{10 K}}$ (the electrical resistivity at 10 K) are converted into the values of $D$ and $\lambda_{tr}$.  This was done using standard values for the electronic coefficient of specific heat for niobium (7.8 mJ/mole/K),\cite{Kittel} an average Fermi velocity for niobium\cite{Kerchner} of 2.7$\times10^{5}$ m/s and using the equations from Orlando {\it et al.}\cite{Orlando}   In addition, using the work of Orlando {\it et al.} and the measured values above, along with known values of the energy gap for niobium, the strong coupling correction term for the upper critical field for these films can be calculated to be $\eta_{\mathrm{Bc2\perp}}\approx 1.07$.  This value will be used to adjust the measured slopes of $T$ vs. $B^{2}$.

\section{Film Structure}

\begin{figure}
\includegraphics*[width=3.0in]{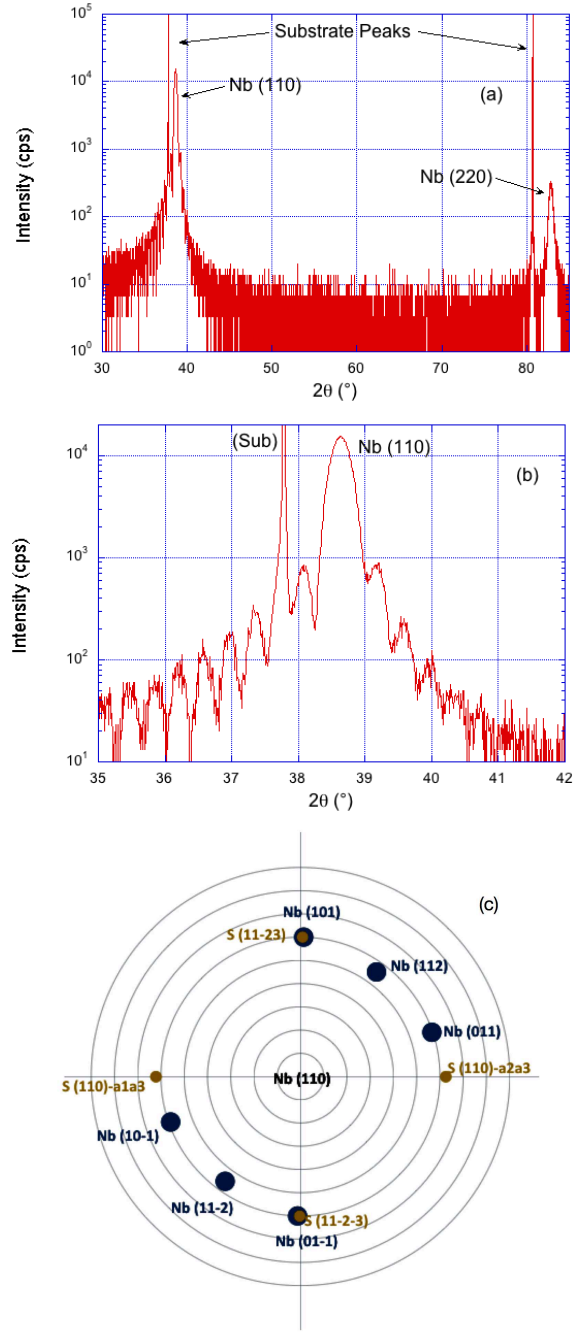}
\caption{\label{XRD} (a) XRD along the film normal for sample H, a 22 nm film on A-plane sapphire, showing only (110) Nb peaks along the growth direction.  The film and substrate peaks are labeled.  (b) Detailed XRD scan along the growth normal for the same niobium film as in (a) around the (110) Nb peak, showing finite thickness fringes.  (c) A ``Pole plot'' for the same niobium film as in (b), formed by combining 4 phi scans for different Bragg angles for the various peaks shown, showing the registry between the niobium film and the substrate.  Here ``S'' stands for the substrate reflections.  Each circle represents a change in angle of 10$^{\circ}$ from the sample normal for the scattering vector.}
\end{figure}

Some samples were characterized by X-ray diffraction, both along the growth direction and in plane, using Cu K$\alpha_{1}$ radiation.  Niobium films grown under the conditions used here (low substrate temperatures) typically exhibit close packed plane (110) growth along the substrate normal with fiber texture (i.e., no in-plane order) .  However, film H with the lowest value of $\lambda_{tr}$=2.4 showed surprising results.  XRD along the growth direction again showed (110) along the growth direction, as seen in Fig. \ref{XRD}a, however the film also showed finite thickness fringes, consistent with the known film thickness of 22 nm, as seen in Fig. \ref{XRD}b.  This sample has a rocking curve width for the (110) reflection of 0.02$^{\circ}$.  In addition, various scans for different values of Bragg angle as a function of angle around the sample normal showed the film does have in-plane texture, as the ``pole plot'' in Fig. \ref{XRD}c demonstrates.  This composite plot shows the result of the various scans, indicating the relationship between the sample reflections and substrate reflections.  The behavior of this film is consistent with that seen in the work by Zhao {\it et al.}\cite{Zhao} on energetically condensed niobium.  The difference here is that this film is grown at a much lower substrate temperature without the additional energetics.  This film was also grown on A-plane sapphire, and it is seen that all such films in this study have the lowest values of $\lambda_{tr}$, or cleaner films.  Although Nb films with very high quality structure have been grown even at room temperature before, those samples are grown on substrates that underwent extreme preparation, including ion milling and UHV annealing to 1200 C.\cite{Claassen}  The substrates here have not had either the benefit of UHV preparation or external energetics, and still produced what appears to be very well ordered Nb films in certain cases.

\section{Upper critical field}

\begin{figure}
\includegraphics*[width=5in]{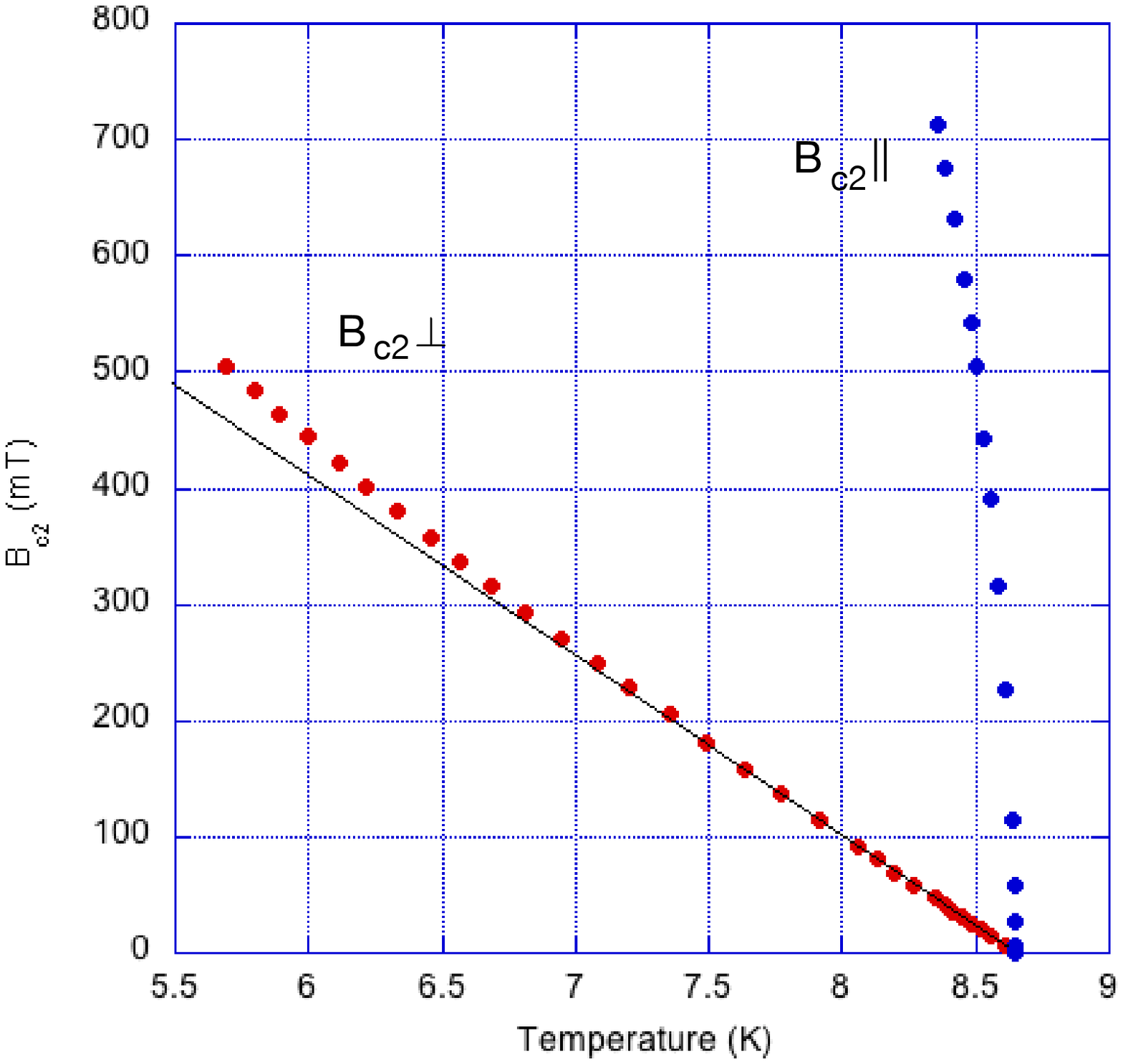}
\caption{\label{para_perp} Perpendicular and parallel upper critical field vs temperature for the 22 nm Nb film H shown in Fig. \ref{XRD} with $\lambda_{\mathrm{tr}}$ = 2.4.   The line on the perpendicular data is a guide to the eye to show the data is linear near $T_{c0}$.} 
\end{figure}

Fig. \ref{para_perp} shows the upper critical field for film H shown in the previous figure with the field perpendicular ($\theta =0^{\circ}$) and parallel ($\theta = 90^{\circ}$) where $\theta$ is the angle between the applied magnetic field and the sample normal.  One can see the expected linear behavior for the perpendicular field near $T_{c0}$, and what looks like the expected square root behavior of the parallel field.  There is a deviation from linearity for the perpendicular field for lower temperatures, but this is believed to be due to effects of anisotropy of the Fermi surface, as seen by Kerchner {\it et al.}\cite{Kerchner} Although this is usually assumed to only be seen in very clean samples ($\lambda_{\mathrm{tr}}<1$), Hohenberg and Werthamer showed this can also be seen in moderately dirty samples.\cite{HW}

\begin{figure}
\includegraphics*[width=5in]{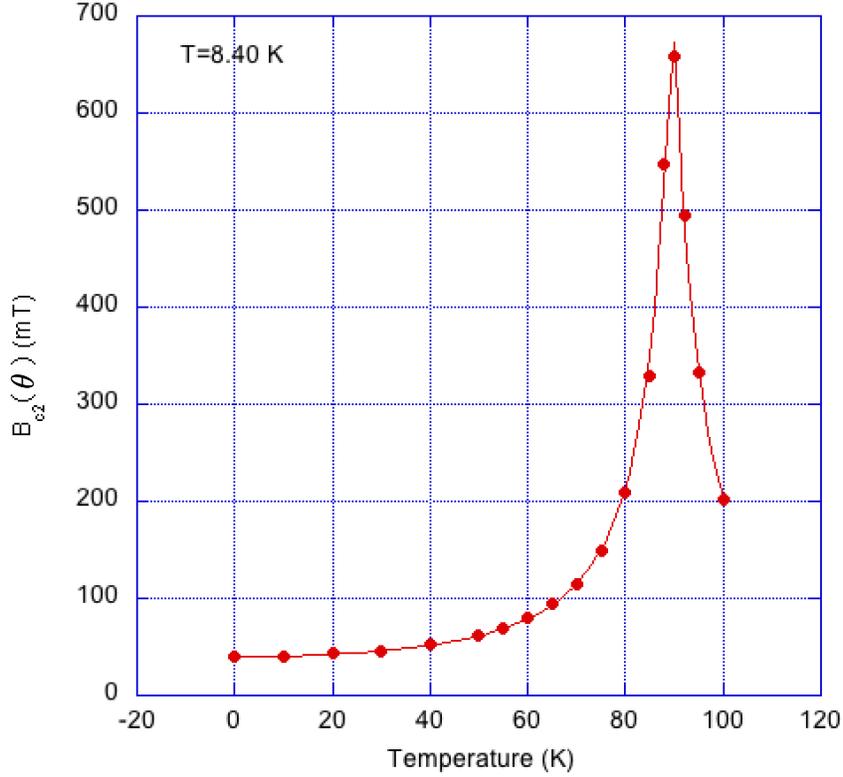}
\caption{\label{2Dfit} Upper critical field vs angle for the 22 nm Nb film H in Fig. \ref{para_perp} at a temperature of 8.40 K, along with the fit to expected 2D behavior from Eq. \ref{Eq6}.}
\end{figure}

\begin{figure}
\includegraphics[width=5in]{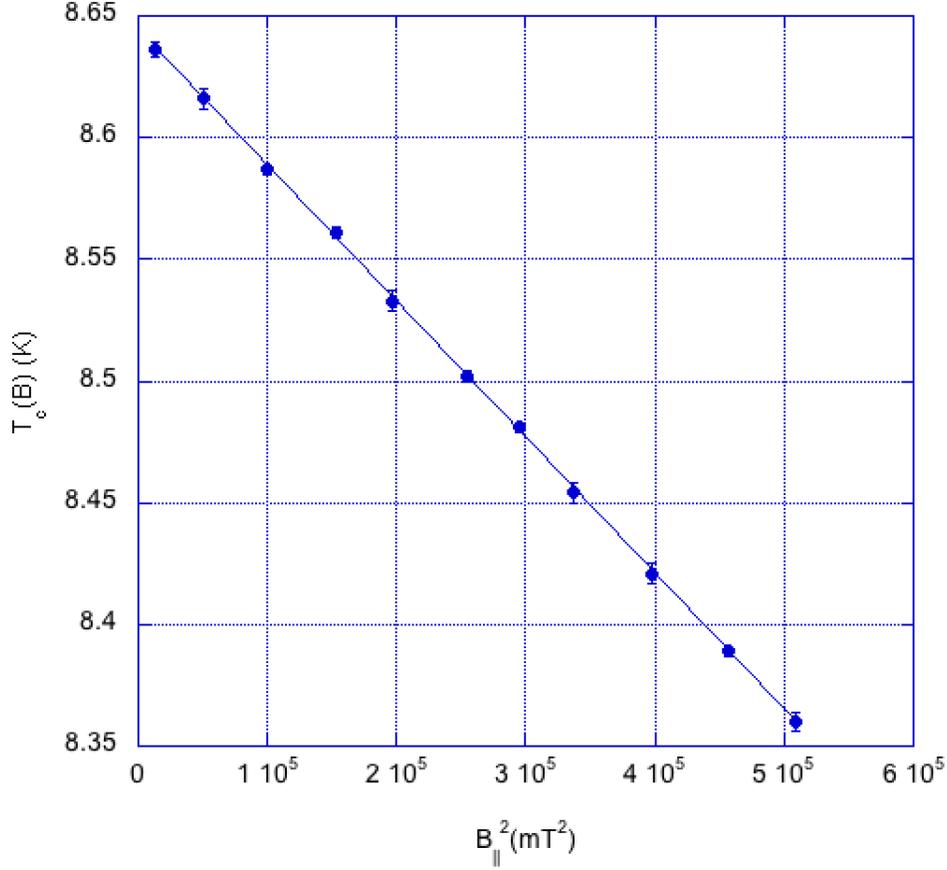}
\caption{\label{para_fit} $T_{c}(B)$ vs $B_{||}^{2}$ for the previous 22 nm Nb film H.  The linear fit accounts for uncertainties in both $T$ and $B$.}
\end{figure}

In Fig. \ref{2Dfit} one can see the angular dependence of the upper critical field for film H at 8.40 K, or $t=0.972$, along with the expected 2D fit, given by Eq. \ref{Eq6}.
  The good fit clearly shows the 2D, or laminar, nature of the superconductivity for the parallel field case for this study.  This angular dependence will hold true for all the models under consideration here.
Fig. \ref{para_fit} shows a plot of $T$ vs $B_{c2||}^{2}$ for the same film shown in the previous figures in order to compare to the equations from Hara and Nagai, Eqs. \ref{Eq3}-\ref{Eq4}.  From this plot the value of $T_{c0}$ can be extracted.  The slope of the curve was then used to extract $C_{2}$ as described in the next section.  This analysis was done for the 16 films used in this study, as shown in Table \ref{Table}.   All films in this study had negative slopes for plots such as this, including two films that each had $d<d_{c}$, films H and L.  These films, both grown on A-plane sapphire, were 22 and 16 nm thick, with $\lambda_{\mathrm{tr}}$ = 2.4 and 5.3, respectively, and had critical thicknesses from Kogan's prediction (Eq. \ref{Eq1}) of 29 and 17 nm, respectively.  Now with thicknesses uncertain to 5\%, the second film, L,  is within uncertainty of its critical thickness.  But even if there is this much deviation, under the Kogan model, the value of $C_{2}$ from Eq. \ref{Eq5} would be either zero or positive.  The first sample, H, is clearly thinner than its critical thickness, even accounting for uncertainty.  Neither sample had a slope of $T$ vs $B_{c2||}^{2}$ that was positive or zero.  It would appear that for these two films, Kogan's analysis is not valid and in the next section, the other films will be considered.

\section{Parallel Critical Field measurements}
The films in this study span a range of  $D=d/\xi$ from 0.37 to 1.4, with $\lambda_{tr}$ varying from 2.4 to 11 (as seen in Table \ref{Table}), which is consistent with the range in the work by Kogan and that of Hara and Nagai.  The analysis for all films was as follows.  First the data was plotted as $T$ vs $B^{2}$, as in Fig. \ref{para_fit} to find the value of $T_{c0}$.  To get the measured value of $C_{2}$, a plot of $t_{C}=T_{c}(B)/T_{c0}$ vs $B^{2}$ was then created. The slope of this plot was then multiplied by $\eta_{\mathrm{Bc2}}$ to remove the strong coupling and then divided by $((e/\hbar)\xi^{2})^{2}$ to give $C_{2}$.  All linear fits were done accounting for uncertainties both in temperature and field using Linefit.\cite{Petcher}

\begin{figure}
\includegraphics[width=5in]{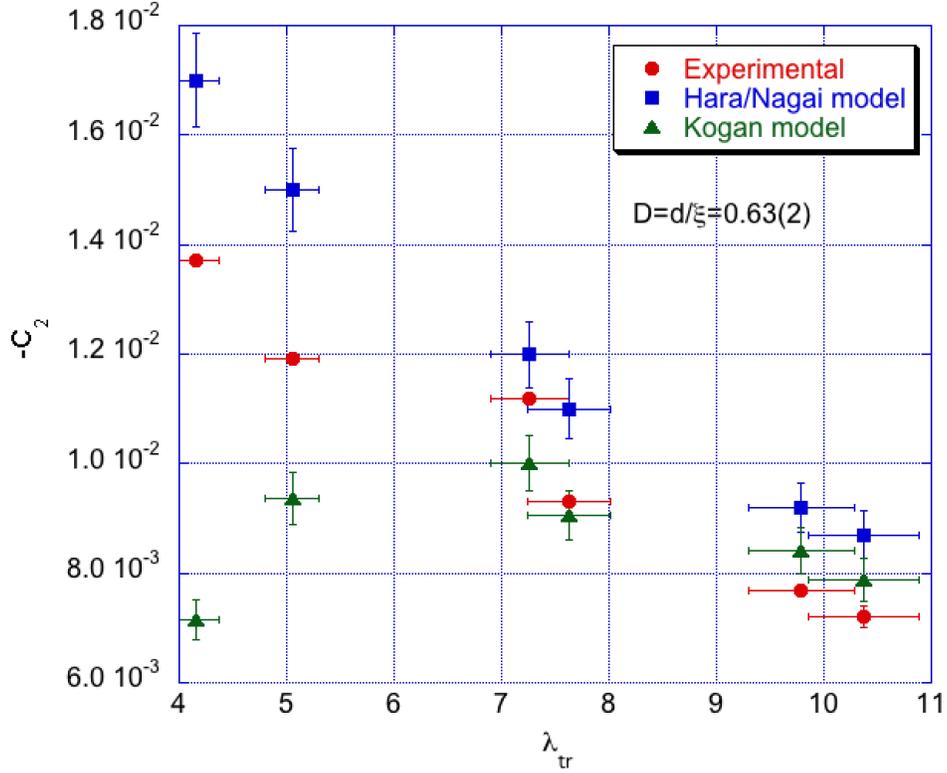}
\caption{\label{C21} Experimental values of -$C_{2}$ vs $\lambda_{\mathrm{tr}}$ for niobium films with $D$ =0.63(2) compared to theoretical values determined from Hara and Nagai's Eq. \ref{Eq4} and Kogan's model as given in Eq. \ref{Eq5}. }
\end{figure}

In Fig. \ref{C21} a comparison between the experimental and theoretical values (both from the model of Hara and Nagai, Eq. \ref{Eq4}, and the model of Kogan, Eq. \ref{Eq5}) of $-C_{2}$ vs $\lambda_{\mathrm{tr}}$ for a series of films that have $D=0.63(2)$ is shown.  A 5\% uncertainty in both the value of $\lambda_{tr}$ and the theoretical values of $-C_{2}$ is included, due to the uncertainty of thickness mentioned earlier.  The uncertainties in the experimental values of $-C_{2}$ are due to the uncertainties due to the linear fitting of $t_{C}$ vs $B^{2}$.  First there is a systematic trend with the experimental values of $-C_{2}$ being lower than those from the model by Hara and Nagai.  Now for larger values of $\lambda_{\mathrm{tr}}$, it appears that the values of $-C_{2}$ from Kogan's model seem to agree with the experimental values better than those from the Hara and Nagai model.  However, as $\lambda_{\mathrm{tr}}$ decreases, there is a larger and larger discrepancy between Kogan's values and the experimental values.  This increasing discrepancy is due to the Kogan model begin able to change sign, so as $\lambda_{\mathrm{tr}}$ decreases, and Kogan's value of $-C_{2}$ begins to head toward zero, the deviation from the experimental values is worse and worse.  Clearly for the data of this figure, the Kogan model cannot fit the experimental results.  It is clear that the model of Hara and Nagai can replicate the overall trend of the experimental data, but continually predicts a larger value than experimental.  

\begin{figure}
\includegraphics[width=6in]{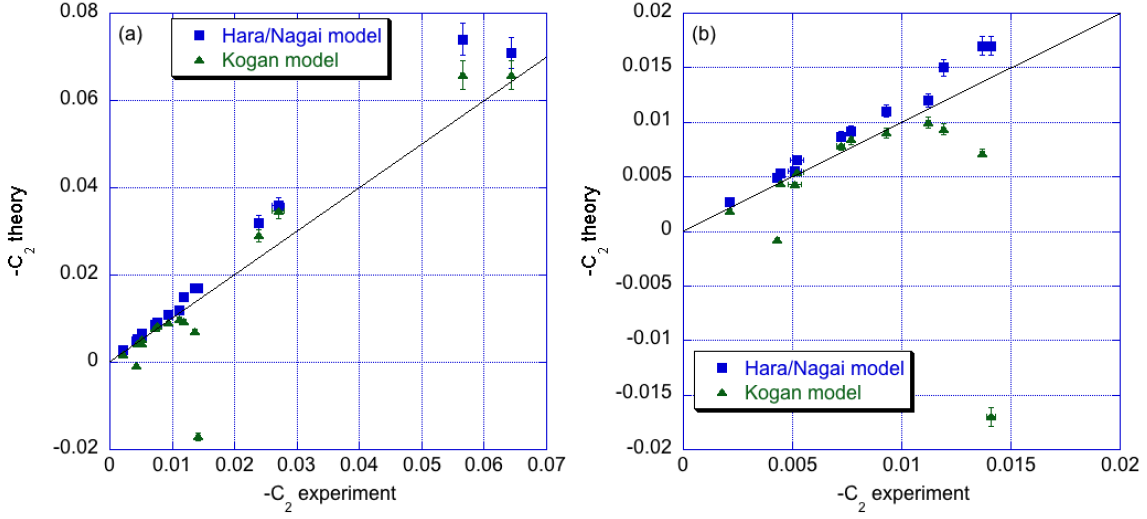}
\caption{\label{C22} (a) Theoretical values of $-C_{2}$ from the Hara-Nagai model as well as Kogan model vs experimental $-C_{2}$ for all the niobium films in this study, and (b) for films with smaller values of $-C_{2}$.  The lines in each plot would indicate exact agreement between experimental and the theoretical values. }
\end{figure}

In Fig. \ref{C22}, a comparison for all samples between the two theoretical models is made against the experimental values. Here the x-axis represents the experimental value of $-C_{2}$, and the y-axis has values for both the Hara and Nagai model and Kogan model.  Clearly for the films with smaller values of $\lambda_{\mathrm{tr}}$, the Kogan values of $-C_{2}$ deviate greatly from the line showing equality between experiment and theory.  For the two samples mentioned earlier, samples H and L, the Kogan values change sign while the experimental values do not.  Again, as stated before, even if sample L is right at the critical thickness instead of below it, it would have a $-C_{2}$ of zero from the Kogan model, which is not consistent with the experimental value.  For the values predicted by the Hara and Nagai model, again there is a systematic trend with experimental values lower than theoretical (by an average value of $\approx$ 20\%), however the values from the Hara and Nagai model do follow the general trend of the experimental values, while for the Kogan values, while there is better agreement for some values, there is wide disagreement for others, typically tied to the samples with smaller values of $\lambda_{\mathrm{tr}}$ and $D$.  

The inability of the Kogan model to fit the data over the entire range again points to a problem with his formulation.  For the Hara and Nagai model, although there is agreement with the overall trend, the continual overestimation of $-C_{2}$ is a puzzlement.  From an experimental standpoint, this implies either measured temperatures are off by 20\% or the magnetic field is off by 10\%, or some combination to give the total.  Neither option seems likely, as the thermometer was calibrated, and the magnetic field was checked by two independent hall probes.  One systematic area in the theory of Hara and Nagai is the integral in Eq. \ref{Eq4}.  This integral is evaluated assuming a spherical Fermi surface, which niobium does not have.  Exactly how to include this in the model is not clear at this time.  However, it is known that non-spherical Fermi surface effects have no impact on the slope of the perpendicular upper critical field near $T_{c0}$.\cite{Kerchner}  Whether they could impact the parallel critical field near $T_{c0}$ under the model of Hara and Nagai is not known.  In addition, in perpendicular critical field, the effect of a non spherical Fermi surface would have less impact as the value of $\lambda_{\mathrm{tr}}$ increases.  For the samples of this study, however, the \% difference observed is approximately the same for samples with $\lambda_{\mathrm{tr}} =11$ or $\lambda_{\mathrm{tr}} =2.4$.  At this point, the exact reason for the systematic deviation between the experimental results and the model of Hara and Nagai is not understood.

\section{Conclusions}
This report has presented the first comparison test between experiment and the models of Kogan and that of Hara and Nagai on parallel critical field slopes near $T_{c0}$.  It is found that there is a systematically lower value of the experimental slope of $t$ vs $B^{2}$ for the niobium films of this study compared to what the Hara and Nagai theory would predict.  There is no evidence of the predicted enhancement of $T_{c}(B)$ by Kogan, and a serious disagreement between the slopes predicted by Kogan and those measured for cleaner, thinner samples.

\section{Acknowledgments}
The author would like to gratefully acknowledge the assistance of A. Hunziker and A. Davis in the film production and measurement, the support of Covenant College for this work, and Prof. C. B. Eom for the X-ray analysis.  This work was performed in part at the Georgia Tech Institute for Electronics and Nanotechnology, a member of the National Nanotechnology Coordinated Infrastructure, which is supported by the National Science Foundation (Grant ECCS-1542174).

\newpage

\section{Appendix}

\begin{table}[!htbp]
\caption{\label{Table}Niobium samples used in this study.  All samples except B and C were grown at elevated temperatures, while B and C were grown at room temperature.  Here $d$ is the sample thickness, Si and Sap denote silicon and sapphire substrates, respectively, and RRR is the residual resistivity ratio between 290 K and 10 K.  Note that uncertainties for $T_{c0}$ are 3 mK, the uncertainties for $-C_{2}$ experimental are given with the values and for all other values uncertainties are as stated in the text.  The last two columns are the theoretical values for $-C_{2}$ for the model of Hara and Nagai (HN) and Kogan.}
\begin{center}
\begin{tabular}{||c|c|c|c|c|c|c|c|c|c|c||}
\hline
Sample & Substrate & $d$ (nm) &RRR & $\rho_{\mathrm{10K}}\ (\mathrm{n}\Omega\mathrm{m}) $& $T_{c0}$ (K)& $D=d/\xi $& $\lambda_{\mathrm{tr}} $& $-C_{2} (\mathrm{Exp.}) $& $-C_{2} (\mathrm{HN}) $& $-C_{2} (\mathrm{Kogan}) $\\
\hline
A & (100) Si & 48 & 4.04 & $49 $& $8.946 $& $1.3 $& $4.4 $& $0.064 $& $0.071 $& $0.066 $\\
\hline
B & (100) Si & 23 & 2.68 & $89 $& $7.777 $& $0.54 $& $10 $& $0.0052(3) $& $0.0065 $& $0.0054 $\\
\hline
C & (100) Si & 48 & 2.68 & $89 $& $8.262 $& $1.2 $& $9.5 $& $0.027(1) $& $0.036 $& $0.035 $\\
\hline
D & (100) Si & 25 & 2.51 & $1.0\times10^{2} $& $8.426 $& $0.63 $& $10 $& $0.0072(2) $& $0.0087 $& $0.0079 $\\
\hline
E & A plane Sap & 25 & 4.82 & $39 $& $8.284 $& $0.63 $& $4.2 $& $0.014 $& $0.017 $& $0.0072 $\\
\hline
F & C plane Sap & 52 & 3.77 & $54 $& $8.649 $& $1.4 $& $5.5 $& $0.056 $& $0.074 $& $0.066 $\\
\hline
G & (100) Si & 25 & 3.13 & $70 $& $8.504 $& $0.65 $& $7.3 $& $0.011 $& $0.012 $& $0.010 $\\
\hline
H & A plane Sap & 22 & 7.24 & $24 $& $8.644 $& $0.57 $& $2.4 $& $0.014 $& $0.017 $& $-0.017 $\\
\hline
J & C plane Sap & 25 & 4.07 & $49 $& $8.477 $& $0.64 $& $5.0 $& $0.012 $& $0.015 $& $0.0094 $\\
\hline
K & C plane Sap & 16 & 2.53 & $98 $& $7.615 $& $0.37 $& $11 $& $0.0021 $& $0.0027 $& $0.0019 $\\
\hline
L & A plane Sap & 16 & 4.12 & $48 $& $8.021 $& $0.39 $& $5.3 $& $0.0043 $& $0.0049 $& $-0.0072 $\\
\hline
M & C plane Sap & 26 & 2.63 & $92 $& $8.264 $& $0.64 $& $9.8 $& $0.0077 $& $0.0092 $& $0.0084 $\\
\hline
N & (100) Si & 24 & 3.01 & $75 $& $8.600 $& $0.63 $& $7.6 $& $0.0093(1) $& $0.011 $& $0.0090 $\\
\hline
P & C plane Sap & 20 & 2.68 & $89 $& $8.050 $& $0.49 $& $9.7 $& $0.0045 $& $0.0053 $& $0.0043 $\\
\hline
R & C plane Sap & 20 & 2.59 & $94 $& $8.060 $& $0.50 $& $10 $& $0.0051(3) $& $0.0055 $& $0.0044 $\\
\hline
S & C plane Sap & 39 & 3.27 & $66 $& $8.520 $& $1.0 $& $6.8 $& $0.024 $& $0.032 $& $0.029 $\\
\hline
\end{tabular}
\end{center}
\label{default}
\end{table}

\end{document}